\documentclass[aps,pra,preprint,groupedaddress]{revtex4-1}

\usepackage{amsmath}   
\usepackage{graphicx}   % for figures

\newcommand{\commentOut}[1]{}

\usepackage[usenames]{color}

\usepackage{times}
\begin{document}

\title{Radiation Damping of a Polarizable Particle}
\author{Lukas Novotny}
 \affiliation{ETH Z{\"u}rich, Photonics Laboratory, 8093 Z{\"u}rich, Switzerland.}
 \email{www.photonics.ethz.ch}   %optional
\date{\today}

\begin{abstract}
A polarizable body moving in an external electromagnetic field will slow down. This effect is referred to as {\em radiation damping} and is analogous to Doppler cooling in atomic physics. Using the principles of special relativity we derive an expression for the radiation damping force and find that it solely depends on the scattered power. The cooling of the particle's center-of-mass motion is balanced by heating due to radiation pressure shot noise, giving rise to an equilibrium that depends on the ratio of the field's frequency and the particle's mass. While damping is of relativistic nature heating has it's roots in quantum mechanics. \\
\end{abstract}

\maketitle

\section{Introduction}
It is well known, that a moving polarizable body that is in thermal equilibrium with its environment will ultimately come to rest~\cite{einstein10,boyer69,milonni94,zurita04}. This frictional force arises from correlations between the fluctuating charges that constitute the body and the external fluctuating fields. It can be shown that these frictional forces originate from blackbody radiation and that zero-point quantum fluctuations have no contribution, in agreement with the Lorentz invariance of motion~\cite{boyer69}. Instead of a thermal field we here consider a particle in an irradiating laser field. \\[-1ex]

Already in 1976,  Ashkin speculated about the damping force of a laser-trapped particle in ultrahigh vacuum~\cite{ashkin76}. This viscous force is of the same nature as the friction experienced by a moving mirror in a radiation field, a configuration first analyzed by Braginsky and co-workers in 1967~\cite{braginski67,braginski70}. This frictional force slows the motion of the mirror down and is referred to as Doppler cooling.  Karrai and co-workers  showed that Doppler cooling depends on the dispersion of the reflecting mirror and that it can be enhanced by orders of magnitude by a photonic crystal~\cite{karrai08}. In atomic physics, the dispersion associated with a two-level system sets limits to the cooling rate, the so-called Doppler limit~\cite{wineland79}.\\[-1ex]

To derive the mechanical force acting on a laser-irradiated dielectric particle we evaluate Maxwell's stress tensor of the scattered radiation and integrate it on an arbitrary enclosing surface.  Similar approaches have been used in other works to derive the radiation reaction force acting on an accelerated charge~\cite{hartemann95,singal17}. These calculations made use of the retarded Lienard-Wichert fields for a charge in arbitrary motion. In principle, an analogous calculation can be performed for an oscillating electric dipole since the retarded fields of a dipole  in motion are known~\cite{power01}. However, in order to avoid the problem of transforming between retarded and proper times we here choose to proceed with the principles of special relativity, making explicit use of the Lorentz transformation to distinguish between the fields in the laboratory frame and the moving particle frame.\\

\section{Problem Statement}
\begin{figure}[!b]
\begin{center}
\includegraphics[width=0.6\textwidth]{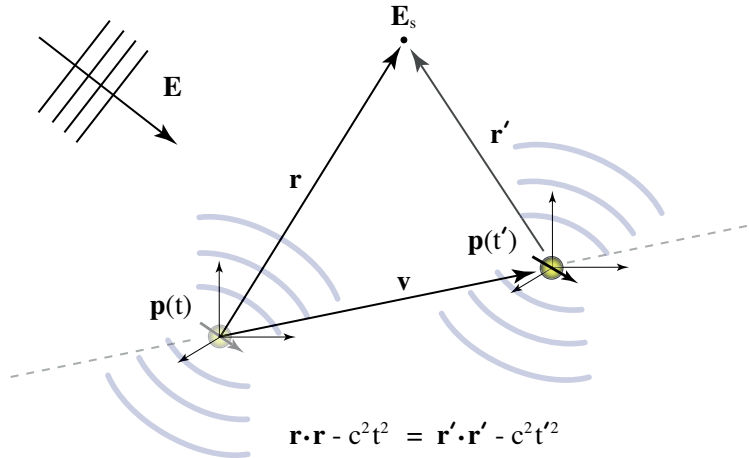}
\end{center}
\caption{A particle moving at velocity ${\bf v}$ is irradiated by the field ${\bf E}$. The induced dipole ${\bf p}$ radiates a field ${\bf E}_s$ that is observed at location ${\bf r}$ and time $t$. In the particle frame, the location and the time are ${\bf r}'$ and $t'$, respectively. 
}
\label{relativity}
\end{figure}

As illustrated in Fig.~\ref{relativity}, we consider a polarizable particle moving with velocity ${\bf v}$. An external field ${\bf E}$ irradiates the particle and induces a dipole ${\bf p}=\alpha{\bf E}$, with $\alpha$ being the particle polarizability. The field ${\bf E}_{\rm s}$ scattered by the particle is recorded at location ${\bf r}$ in the laboratory frame. The same point measured from the particle frame is ${\bf r}'$. \\

%
%\subsection{Special Relativity}
The postulates of special relativity require that the velocity of the field traveling from the particle to the observation point is independent of the reference frame, that is,
\begin{equation}
c^2\; =\; \frac{{\bf r}\cdot{\bf r}}{t^2}\;=\;\frac{{\bf r}'\cdot{\bf r}'}{t'^2} \; ,
\label{postul}
\end{equation}
where $t$ and $t'$ are the times in the laboratory frame and the particle frame, respectively. We express the incident fields ${\bf E}$ and ${\bf B}$ in terms of their projections parallel and perpendicular to ${\bf v}$, that is, ${\bf E} = {\bf E}_{\parallel} + {\bf E}_{\perp}$ with ${\bf E}_{\parallel}=({\bf E}\cdot{\bf v}) {\bf v} /v^2$ and  ${\bf E}_{\perp}=({\bf E}\times{\bf v}) /v$, and similarly for the magnetic field ${\bf B}$. In the rest frame of the particle the incident fields are~\cite{zangwill12}
\begin{eqnarray}
{\bf E}_{\parallel}' \,=\, {\bf E}_{\parallel}\quad && 
{\bf E}_{\perp}' \,=\, \gamma \left[{\bf E} + \mbox{\boldmath{$\beta$}}\times c {\bf B}\right]_{\perp}  \label{fieldtrf} \\
{\bf B}_{\parallel}' \,=\, {\bf B}_{\parallel}\quad && 
{\bf B}_{\perp}' \,=\, \gamma \left[{\bf B} - \mbox{\boldmath{$\beta$}}\times {\bf E}/c\right]_{\perp}  , \nonumber
\end{eqnarray}
where $\mbox{\boldmath{$\beta$}}={\bf v}/c$ and $\gamma = 1/\sqrt{1-v^2/c^2}$. \\

In the laboratory frame, space and time are represented by the four-vector $({\bf r}, i c t)$. On the other hand, in the rest frame of the particle, the four-vector is $({\bf r}', i c t')$. The two four-vectors are related by the Lorentz transform, which yields
\begin{eqnarray}
{\bf r}' & = & {\bf r}\,+\, {\bf v}\left[
\frac{{\bf r}\cdot{\bf v}}{v^2} (\gamma-1) \,-\, \gamma\:\!t\right] \\[0.8ex]
t' &=& \gamma\;\!  \left[t \,-\, \frac{{\bf r}\cdot{\bf v} }{c^2}\right]\; .
\end{eqnarray}
The transformed coordinates satisfy condition (\ref{postul}). In terms of the projections of ${\bf r}$ parallel and perpendicular to ${\bf v}$ these relations can be rewritten as
\begin{eqnarray}
{\bf r}'_{\perp} \,=\, {\bf r}_{\perp} \, ,\quad 
{\bf r}'_{\parallel} \,=\, \gamma\:\! ({\bf r}_{\parallel} - \mbox{\boldmath{$\beta$}}\:\! c\:\! t) \, , \quad
t' \,=\, \gamma\;\! (t - \mbox{\boldmath{$\beta$}} \cdot {\bf r}_{\parallel}/c)\; .\\[0.1ex] \nonumber
\end{eqnarray}

\section{Excitation by a Plane Wave}
We now assume that the incident field corresponds to a plane wave polarized in $x$ direction and propagating in $z$ direction. In the laboratory frame it reads as
\begin{equation}
{\bf E}_{\rm i}(x,y,z,t) \:=\: {\rm Re}\left\{ E_0 \, 
{\rm e}^{i\:\![k\:\!z- \omega t]}\right\}\, {\bf n}_{\rm x} \; ,
\label{plw01}
\end{equation}
with $k=\omega/c$ being the wavevector and ${\bf n}_{\rm x}$ the unit vector in $x$ direction. For a particle moving in the $y$ direction the space-time four-vector in the particle frame is\footnote{The inverse transform is $(x,y,z,ict) \,=\, (x',\, \gamma\:\! [y' + v\:\! t'],\, z',\, i c \gamma\;\! [t' + y' v / c^2])$.}
\begin{eqnarray}
(x',y',z',ict') \,=\, (x,\, \gamma\:\! [y - v\:\! t],\, z,\, i c \gamma\;\! [t - y v / c^2])
\label{lorentz01}
\end{eqnarray}
and the field strength is (c.f. Eqs.~\ref{fieldtrf})
\begin{eqnarray}
E_0' \,=\, \gamma E_0\; .
\end{eqnarray}
Thus, the field felt by the particle in its own reference frame is
\begin{equation}
{\bf E}_{\rm i}'(x',y',z',t') \:=\: {\rm Re}\left\{ \gamma E_0 \,{\rm e}^{i\:\![k\:\!z'- \omega \gamma\;\! (t + y' v / c^2)]} \right\}\, {\bf n}_{\rm x} \; =\;
{\rm Re}\left\{ \gamma E_0 \,{\rm e}^{i\:\![k\:\!z'- k \beta \gamma\;\! y'  - \omega \gamma\;\! t']}\right\}\, {\bf n}_{\rm x} \; .
\label{plw02}
\end{equation}
Here, the four-vector $({\bf k}, i\omega/c)= (0,0,k,i\omega/c)$ gets transformed to $({\bf k}', i\omega'/c)= (0,-k \beta \gamma,k,i\omega\gamma/c)$. Note that  $({\bf k}'\!\cdot {\bf r}' - \omega' t')=({\bf k}\cdot {\bf r} - \omega t)$ is Lorentz invariant and that $\nabla'\cdot{\bf E}_{\rm i}'$ is  zero.\\

The field ${\bf E}'_{\rm i}$ polarizes the particle and induces a dipole ${\bf p}(\omega') = \alpha(\omega') \:\!{\bf E}'_{\rm i}(\omega')$, which in turn radiates a field\footnote{We assume that the particle is observed from large distance and only account for the farfield.}~\cite{novotny12}
 \begin{equation}
{\bf E}'_{\rm s}(x',y',z',t') \:=\: {\rm Re}\left\{ \omega'^2\mu_0\,p(\omega')\, \frac{\exp[i (k' r' - \omega' t')]}{4\pi r'^3} \left[\!
\begin{array}{c}
 y'^2\!+\!z'^2 \\
\!\!-x' y'  \\
\!\!-x' z'
\end{array}
\!\right] \right\}\; ,
\label{plw03a}
\end{equation}
where $x'^2+y'^2+z'^2=r'^2$. The corresponding magnetic field is
 \begin{equation}
{\bf B}'_{\rm s}(x',y',z',t') \:=\: \frac{1}{c}{\rm Re}\left\{ \omega'^2\mu_0\,p(\omega')\, \frac{\exp[i (k' r' - \omega' t')]}{4\pi r'^2} \left[\!
\begin{array}{c}
\,0 \\
\,z'  \\
\!-y'
\end{array}
\!\right] \right\}\; .
\label{plw03b}
\end{equation}
According to (\ref{fieldtrf}), in the laboratory frame these fields  become 
 \begin{equation}
{\bf E}_{\rm s} \:=\:
\left[\! \begin{array}{c} 
\gamma (E'_{{\rm s}_x}  - \beta c B'_{{\rm s}_z}) \\
E'_{{\rm s}_y}  \\
\gamma E'_{{\rm s}_z} 
\end{array}
\!\right]
\quad {\rm and}\quad
{\bf B}_{\rm s} \:=\:
\left[\! \begin{array}{c} 
\gamma \beta E'_{{\rm s}_z} / c \\
B'_{{\rm s}_y}  \\
\gamma (B'_{{\rm s}_z} - \beta E'_{{\rm s}_x}/c) 
\end{array}
\!\right] \; ,
\label{fieldtransf}
\end{equation}
which gives
 \begin{equation}
{\bf E}_{\rm s}(x',y',z',t') \:=\: {\rm Re}\left\{ \omega'^2\mu_0\,p(\omega')\, \frac{\exp[i (k' r' - \omega' t')]}{4\pi r'^3} \left[\!
\begin{array}{c}
 \gamma ( y'^2+z'^2+\beta y' r') \\
\!\!-x' y'  \\
\!\!-\gamma x' z'
\end{array}
\!\right] \right\}\; ,
\label{plw07}
\end{equation}
% \begin{equation}
%{\bf B}_{\rm s}(x',y',z',t') \:=\: \frac{1}{c}{\rm Re}\left\{ \omega'^2\mu_0\,p(\omega')\, \frac{\exp[i (k' r' - \omega' t')]}{4\pi r'^3} \left[\!
%\begin{array}{c}
%-\gamma\beta x'z'\\
%z' r' \\
%\!-\gamma (y' r'+ \beta y'^2+ \beta z'^2)
%\end{array}
%\!\right] \right\}\; .
%\label{plw08}
%\end{equation}
with $k' = k\sqrt{1+\beta^2\gamma^2}= \gamma k$, $\omega'=\gamma\omega$, and $p(\omega')= \alpha(\omega') \gamma E_0 \exp[i\:\!(k'\:\!z'/\gamma- k' \beta\;\! y'  - \omega' t')]$.\\

Moving back into the coordinates of the laboratory frame we have
 \begin{equation}
{\bf E}_{\rm s}(x,y,z,t) \:=\: {\rm Re}\left\{ \omega'^2\mu_0\,\gamma\,p(\omega')\, \frac{\exp[i (k' r' - \omega' t')]}{4\pi r'^3} \left[\!
\begin{array}{c}
 \gamma^2\:\! [y - v\:\! t]^2+z^2+\beta \gamma\:\! [y - v\:\! t] r' \\
\!\!-x\:\! [y - v\:\! t]  \\
\!\!- x z
\end{array}
\!\right] \right\}\; ,
\label{plw10}
\end{equation}
with $r' = \sqrt{r^2 + \gamma^2 [y - v\:\! t]^2 - y^2}$ and $t' = \gamma\;\! [t - y v / c^2]$. Introducing the instantaneous position vector ${\bf R} = {\bf r} \!-\! {\bf v} t = [ x,\, y\!-\! vt,\,z]^T$
and the angle ${\bf n}_v \cdot {\bf n}_R = \cos\theta$ we can express the distance $r'$ as
\begin{eqnarray}
r'^2 &=& x^2 +   [y - v\:\! t]^2 + z^2 + (\gamma^2-1) [y - v\:\! t]^2 \\
&=& R^2 + (\gamma^2-1) [y - v\:\! t]^2 \;=\; R^2 + (\gamma^2-1) [{\bf R}\cdot{\bf n}_v]^2 \nonumber \\
&=& R^2 + (\gamma^2-1) R^2 \cos^2\!\theta \; =\; R^2 (\sin^2\!\theta + \gamma^2\cos^2\!\theta) \nonumber
\end{eqnarray}
and the scattered field becomes (using $y-v t = R\cos\theta$, $x= R\sin\theta\sin\phi$, $z = R\sin\theta\cos\phi$)
 \begin{eqnarray}
{\bf E}_{\rm s}(x,y,z,t) &=& {\rm Re}
%\left
{\Bigg \{} \mu_0\:\!\omega^2\gamma^3 p(\gamma\omega)\, \frac{\exp[i \gamma (k r - \omega t)]}{4\pi R (\sin^2\!\theta + \gamma^2\cos^2\!\theta)^{3/2}} 
%\right. 
\;\times \label{plw11}
 \\[1ex] 
&&\qquad\qquad \left.
\left[\!
\begin{array}{c}
  ( \gamma^2\:\! \cos^2\!\theta +\sin^2\!\theta \cos^2\phi+\beta \gamma\:\!  \cos\theta \sqrt{\sin^2\!\theta + \gamma^2\cos^2\!\theta}) \\
\!\!-  \sin\theta \cos\theta \sin\phi \\
\!\!- \sin^2\!\theta \cos\phi \sin\phi
\end{array}
\!\right] \right\}\; .
\nonumber
\end{eqnarray}
Note that the wavefronts $\exp[i (k' r' - \omega' t')]$  are no longer spherical in the laboratory frame. However, since $\gamma=1+ \beta^2/2+3 \beta^4/8+ ..$ the deviation from a spherical surface is second-order in $\beta$ and hence, to first-order, we can treat them as spherical.\\

\section{Force Acting on the Particle}
We now calculate the mean force acting on the particle using Maxwell's stress tensor 
{\color{black} 
$\tensor{\bf T}\,=\varepsilon_0 {\bf E}{\bf E} + \mu_0  \; {\bf H}{\bf H} - \frac{1}{2} (\varepsilon_0 {\bf E}\cdot{\bf E} + \mu_0  \; {\bf H}\cdot{\bf H}) \tensor{\bf I}$
 as~\cite{novotny12,zangwill12}
\begin{equation}
\left\langle {\bf F}\right\rangle \;=\; \int_{\partial V} \langle \tensor{ \bf T}({\bf r},t)\rangle \cdot {\bf n}'\;da \, ,
   \label{tensor05}
\end{equation}
where the brackets $\langle\, .. \, \rangle$ denote the time average. The fields ${\bf E}$ and ${\bf H}$ are the sum of incident and scattered fields, that is, ${\bf E}={\bf E}_{\rm i}+{\bf E}_{\rm s}$ and ${\bf H}={\bf H}_{\rm i}+{\bf H}_{\rm s}$, respectively. Thus, there are three different contributions to $\left\langle {\bf F}\right\rangle$, one related to ${\bf E}_{\rm i}$, one to ${\bf E}_{\rm s}$, and one to the interference between ${\bf E}_{\rm i}$ and ${\bf E}_{\rm s}$
\begin{equation}
\left\langle {\bf F}\right\rangle \;=\; \left\langle {\bf F}_{\rm i}\right\rangle + \left\langle {\bf F}_{\rm s}\right\rangle + \left\langle {\bf F}_{\rm is}\right\rangle \, .
   \label{tensor06}
\end{equation}
It is straightforward to show that $\left\langle {\bf F}_{\rm i}\right\rangle=0$. Furthermore, to lowest order in $\beta$ the interference term yields~\footnote{The derivation of this result requires to include also the near-field ($\propto R^{-3}$) and the intermediate field  ($\propto R^{-2}$) of ${\bf E}_{\rm s}$.}
\begin{equation}
\left\langle {\bf F}_{\rm is}\right\rangle\; =\;  \frac{1}{2}\;\! {\rm Im}\{\alpha\}\,k\,E_0^2\,{\bf n}_{\rm z}\;=\; \frac{\sigma}{c}\, I_0\,{\bf n}_{\rm z}\;=\; \frac{1}{c} P_{\rm scatt} \,{\bf n}_{\rm z}\; ,
   \label{tensor06y}
\end{equation}
   where we introduced the scattering cross-section $\sigma = {\rm Im}\{\alpha\}\ k /\varepsilon_0$ (assuming no intrinsic absorption), the intensity
 $I_0=(1/2) \varepsilon_0 c E_0^2$, and the scattered power $P_{\rm scatt} =\sigma\:\!I_0$. Eq.~(\ref{tensor06y}) is just the standard expression for the radiation pressure force acting on a polarizable object in direction of wave propagation. Since it is independent of $\beta$ there is no friction associated with this term.\\

To calculate the contribution of $\left\langle {\bf F}_{\rm s}\right\rangle$ } we choose an enclosing surface $\partial V$ that coincides with the wavefronts such that ${\bf n}'$ is perpendicular to ${\bf E}_{\rm s}$ everywhere on the surface. In this case, (\ref{tensor05}) can be written as
\begin{equation}
\left\langle {\bf F_{\rm s}}\right\rangle \;=\; -\varepsilon_0 \int_{\partial V}  
 \langle {\bf E}_{\rm s}\cdot{\bf E}_{\rm s} \rangle\,  {\bf n}'\;da \, ,
   \label{tensor06x}
\end{equation}
where we used the fact that in the far-zone  the energy density of the magnetic field is the same as the energy density of the electric field. In the limit $v\ll c$ the enclosing surface can be approximated by a spherical surface, such that ${\bf n}'\approx{\bf n}=[\sin\theta\sin\phi,\,\cos\theta,\,\sin\theta\cos\phi]^T$. Then, using (\ref{plw11}) in (\ref{tensor06x}) we obtain
\begin{equation}
\left\langle {\bf F_{\rm s}}\right\rangle \;=\; - {\bf n}_{\rm y}\,\frac{\varepsilon_0}{2} \frac{2\pi\mu_0^2 \omega^4 \gamma^6}{16\pi^2} |p(\omega')|^2  \beta\gamma
 \int_{0}^{\pi}  \frac{\cos\theta [2\gamma^2\cos^2\!\theta+\sin^2\!\theta]}{[\gamma^2\cos^2\!\theta+\sin^2\!\theta]^{5/2}}\,\sin\theta\,d\theta\, ,
   \label{tensor07}
\end{equation}
where we carried out the integration over $\phi$ and dropped all terms that cancel upon integration over $\theta$. The integral in (\ref{tensor07}) can be evaluated and, to leading order in $v/c$, yields a value of $16/15$. Note that the leading order in the expression for $\left\langle {\bf F_{\rm s}}\right\rangle$ is defined by $\beta$ and that $\gamma$, and powers thereof, contribute only to second order in $v/c$. The resulting force turns out to be
\begin{equation}
\left\langle {\bf F_{\rm s}}\right\rangle \;=\; -\frac{4}{5} \frac{P_{\rm scatt}(\omega')}{c^2} v\,{\bf n}_{\rm y}\, .     \label{tensor08}
\end{equation}
where $P_{\rm scatt}(\omega') =  |p(\omega')|^2 {\omega'}^4 / (12\pi c^3 \varepsilon_0)$ is the scattered power in the moving particle frame. Thus, it turns out that the force is proportional to the particle velocity and therefore constitutes a frictional term, which we denote as {\em radiation damping}. Our calculation shows that it is of purely relativistic nature and originates from the interference between the magnetic and the electric terms in the expression of the scattered field ${\bf E}_{\rm s}$ in (\ref{fieldtransf}).\\

\section{Arbitrary Particle Motion}
In the calculation above we have considered a particle moving in $y$ direction and the calculation can be repeated for motions in $x$ and $z$ directions (c.f. Appendix A and B). We find that for a particle moving in the direction of wave propagation ($z$ direction) the damping force is the same as in the case considered above ($y$ direction). On the other hand, for a particle moving in the direction of polarization ($x$ direction) the damping is only half as much. Combining the results for the three directions we obtain 
\begin{equation}
\, \left\langle {\bf F}_{\rm s} \right\rangle \;=\; -\,\frac{P_{\rm scatt}(\omega')}{5\,c^2}\, \left[\!
\begin{array}{ccc}
2 & 0 & 0\\[-0.5ex]
0 & 4 & 0\\[-0.5ex]
0 & 0 & 4\
\end{array}
\! \right]\, \dot{\bf r} \; ,
   \label{totforcefric}
\end{equation}
where the argument of $P_{\rm scatt}$ indicates that the scattered power is to be evaluated in the particle frame. To lowest order in $v/c$ we find $P_{\rm scatt}(\omega')=P_{\rm scatt}(\omega)$. From a photon picture of view, radiation damping is equivalent to Doppler cooling\cite{wineland79}. It slows the particle motion down.\\

\section{Equilibrium in the Radiation Field}
From the discussion above it follows that a particle moving in an external electromagnetic field will slow down due to radiation damping, that is, relativity cools the particle's center-of-mass motion. On the other hand, the particle's motion is heated up by the momentum transfer of photons that scatter off the particle~\cite{chang10,jain16a}. This radiation pressure shot-noise is of quantum nature as it relies on the discreteness of the radiation field. Thus, while relativity slows the particle motion quantum mechanics accelerates it! The balance between heating and cooling defines the particle's equilibrium energy, which can be expressed in terms of an effective center-of-mass temperature $T$.\\[-1ex]

Let us consider a particle moving in $y$ direction. It's equation of motion is
\begin{equation}
\ddot y \,+\, \Gamma\:\! \dot y \;=\; \frac{1}{m}\, F_{\rm fluct}(t) \; ,
\end{equation}
where $\Gamma = (4/5)  P_{\rm scatt}\,/\,mc^2$ (c.f. Eq.~\ref{totforcefric}). The fluctuating force $F_{\rm fluct}$ is associated with shot noise, that is, with the discreteness of the radiation field. Its power spectral density is~\cite{jain16a}
\begin{eqnarray}
S_{F}(\Omega)\;=\; \frac{1}{2\pi} \int_{-\infty}^{\infty} \left\langle F_{\rm fluct}(t)\,F_{\rm fluct}(t+t')\right\rangle\, {\rm e}^{i\Omega t'} dt'  \;=\; \frac{2}{5}  \frac{\hbar\omega}{2\pi c^2} \, {P}_{{\rm scatt}} \; ,
\end{eqnarray}
where $\omega$ is the angular frequency of the laser field. In the steady-state there is a balance between damping ($\Gamma$) and heating ($S_F$), such that the average energy  of the particle remains constant. This balance is formulated in terms of the fluctuation-dissipation theorem~\cite{kubo85}
\begin{eqnarray}
S_F  \;=\;  \frac{m\, k_B T}{\pi} \, \Gamma\; .
\label{fldpth}
\end{eqnarray}
Inserting the expressions for $S_F$ and $\Gamma$ we obtain the center-of-mass temperature of the particle
\begin{equation}
T \;=\;  \frac{1}{4} \frac{\hbar\omega}{k_B}  \,
\label{sttemp}
\end{equation}
with identical expressions for the $x$ and $z$ axes.  In the absence of any restoring forces, the equilibrium velocity of the particle becomes
\begin{equation}
v \;=\;  \frac{1}{2} \sqrt{\frac{\hbar\omega}{m}}  \, .
\label{vstead}
\end{equation}
which follows from the expression of the kinetic energy $m v^2/2$.  For a silica nanoparticle of size $100\,$nm and mass density $\rho_{\rm SiO_2} = 2200\,$kg/m$^3$ irradiated by a laser of wavelength $\lambda=1064\,$nm the steady-state velocity is $v=0.2\,$m/s. Interestingly, the steady-state is independent of the power of the radiation field. This is because both photon recoil heating and radiation damping depend on the scattered power. However, the lower the power is the longer it takes to reach the steady-state. \\

\section{Conclusions}
Using the principles of special relativity we have derived an expression for the radiation damping of a polarizable particle moving in an external radiation field. The damping force scales linearly with the scattered power and depends on the direction of motion relative to the polarization of the field. Since the same directional dependence exists for radiation pressure shot noise the steady-state energy turns out to be independent of direction. We find an equilibrium velocity that depends on the ratio of radiation frequency $\omega$ and the mass $m$ of the particle. It is interesting to note that this equilibrium is guaranteed by the interplay of relativity (cooling) and quantum mechanics (heating). \\

\begin{acknowledgments}
The author thanks  Martin Frimmer for valuable discussions.  The work has been supported by ETH Z{\"u}rich and  ERC-QMES (No. 338763).
\end{acknowledgments}

\newpage

\section*{Appendix A: Movement in $x$ direction}
We now repeat the calculation for a particle moving in $x$ direction, that is, in direction of the polarization ${\bf E}_0$. The field strength in the particle frame is (c.f. Eqs.~\ref{fieldtrf})
\begin{eqnarray}
E_0' \,=\,  E_0\; .
\end{eqnarray}
and the scattered fields (\ref{plw03a}) and (\ref{plw03b}) remain of the same form. According to (\ref{fieldtrf}), in the laboratory  frame these fields become 
 \begin{equation}
{\bf E}_{\rm s} \:=\:
\left[\! \begin{array}{c} 
E'_{{\rm s}_x}  \\
\gamma (E'_{{\rm s}_y} + \beta c B'_{{\rm s}_z})  \\
\gamma (E'_{{\rm s}_z}  - \beta c B'_{{\rm s}_y})
\end{array}
\!\right]\, .
\label{fieldtransfx}
\end{equation}
Moving back into the coordinates of the laboratory frame we have
 \begin{equation}
{\bf E}_{\rm s}(x,y,z,t) \:=\: {\rm Re}\left\{ \omega'^2\mu_0\,p(\omega')\, \frac{\exp[i (k' r' - \omega' t')]}{4\pi r'^3} \left[\!
\begin{array}{c}
y^2 + z^2\\
 -\gamma [\gamma (x-v t) y + \beta y r'] \\
 -\gamma [\gamma (x-v t) z + \beta z r'] 
\end{array}
\!\right] \right\}\; ,
\label{plw10b}
\end{equation}
with $r' = \sqrt{r^2 + \gamma^2 [x - v\:\! t]^2 - x^2}$ and $t' = \gamma\;\! [t - x v / c^2]$. Using $x-v t = R\cos\theta$, $y= R\sin\theta\cos\phi$, $z = R\sin\theta\sin\phi$ yields
\begin{equation}
\left\langle {\bf F}_{\rm s}\right\rangle \;=\; {\bf n}_{\rm x}\,\frac{\varepsilon_0}{2} 
\frac{4\pi\mu_0^2 \omega^4 \gamma^6}{16\pi^2} |p(\gamma\omega)|^2  \beta\gamma
 \int_{0}^{\pi}  \frac{\sin^2\!\theta \cos^2\theta}{[\gamma^2\cos^2\!\theta+\sin^2\!\theta]^{5/2}}\,\sin\theta\,d\theta\, ,
   \label{tensor07b}
\end{equation}
As before, the integral can be integrated and to lowest order yields a value of $4/15$. Thus, to lowest order in $v$
\begin{equation}
\left\langle {\bf F}_{\rm s}\right\rangle  \;=\; -\frac{2}{5} \frac{P_{\rm scatt}(\omega')}{c^2} v\,{\bf n}_{\rm x}\, ,   
\label{tensor08b}
\end{equation}
which is half the force derived for the particle moving in $y$ direction (c.f. Eq.~\ref{tensor08}). Note that because $E_0'=E_0$ the scattered power for a particle moving in $x$ direction is a factor of $1/\gamma^2$ lower than for a particle moving in $y$ direction. \\

\section*{Appendix B: Movement in $z$ direction}
For a particle moving in $z$ direction the field strength in the particle frame becomes (c.f. Eqs.~\ref{fieldtrf})
\begin{eqnarray}
E_0' \,=\,  \gamma [E_0 - \beta c B_0] \,=\,\gamma (1-\beta) E_0\, .
\end{eqnarray}
Using (\ref{fieldtrf}) the amplitude of the scattered field (\ref{plw03a}) in the laboratory frame is
 \begin{equation}
{\bf E}_{\rm s} \:=\:
\left[\! \begin{array}{c} 
\gamma (E'_{{\rm s}_x} + \beta c B'_{{\rm s}_y})  \\
\gamma E'_{{\rm s}_y}  \\
E'_{{\rm s}_z} 
\end{array}
\!\right]\, .
\label{fieldtransfz}
\end{equation}
Moving back into the coordinates of the laboratory frame we have
 \begin{equation}
{\bf E}_{\rm s}(x,y,z,t) \:=\: {\rm Re}\left\{ \omega'^2\mu_0\,p(\omega')\, \frac{\exp[i (k' r' - \omega' t')]}{4\pi r'^3} \left[\!
\begin{array}{c}
\gamma (y^2 + \gamma^2 (z-v t)^2 + \beta \gamma (z-v t) r'\\
 -\gamma x y \\
 -\gamma x  (z-v t)
\end{array}
\!\right] \right\}\; ,
\label{plw10c}
\end{equation}
with $r' = \sqrt{r^2 + \gamma^2 [z - v\:\! t]^2 - z^2}$ and $t' = \gamma\;\! [t - z v / c^2]$. Using $z-v t = R\cos\theta$, $x= R\sin\theta\cos\phi$, $y = R\sin\theta\sin\phi$ yields
\begin{equation}
\left\langle {\bf F}_{\rm s}\right\rangle \;=\; -{\bf n}_{\rm z}\,\frac{\varepsilon_0}{2} 
\frac{2\pi\mu_0^2 \omega^4 \gamma^6}{16\pi^2} |p(\gamma\omega)|^2  \beta\gamma
 \int_{0}^{\pi}  \frac{\cos^2\!\theta [2 \gamma^2\cos^2\!\theta+\sin^2\!\theta]}{[\gamma^2\cos^2\!\theta+\sin^2\!\theta]^{5/2}}\,\sin\theta\,d\theta\, ,
   \label{tensor07c}
\end{equation}
The integral can be evaluated and to lowest order yields a value of $16/15$.  Thus, to lowest order in $v$
\begin{equation}
\left\langle {\bf F}_{\rm s}\right\rangle  \;=\; -\frac{4}{5} \frac{P_{\rm scatt}(\omega')}{c^2} v\,{\bf n}_{\rm z}\, ,   
\label{tensor08c}
\end{equation}
which is the same as the force derived for the particle moving in $y$ direction (c.f. Eq.~\ref{tensor08}). Note that because $E_0'=\gamma (1-\beta) E_0$ the scattered power for a particle moving in $z$ direction is a factor of $(1-\beta)^2\approx(1-2\beta)$ lower than for a particle moving in $y$ direction. \\

\bibliographystyle{apsrev4-1}
%\bibliography{damping.bib}

%

\end{document}